\documentclass[12pt]{iopart}

\usepackage{epsf}
\usepackage{graphicx}

\begin{document}

\title[Anomalous First Order Transition in NSMO]{Anomalous First Order Transition in $Nd{_{0.5}}Sr{_{0.5}}MnO{_{3}}$: An interplay between kinetic arrest and thermodynamic transitions}

\author{R Rawat, K Mukherjee, Kranti Kumar, A Banerjee and P Chaddah}

\address{UGC-DAE Consortium for Scientific Research\\University Campus, Khandwa Road\\
Indore-452017, M.P, India.}
\ead{alok@csr.ernet.in (A Banerjee)}
\begin{abstract}
A detailed investigation of the first order antiferromagnetic insulator (AFI) to ferromagnetic metal (FMM) transition in $Nd{_{0.5}}Sr{_{0.5}}MnO{_{3}}$ is carried out by resistivity and magnetization measurements. These studies reveal several anomalous features of thermomagnetic irreversibility across the first order transition. We show that these anomalous features can not be explained in terms of supercooling effect alone and H-T diagram based on isothermal MH or RH measurement alone do not reflect true nature of the first order transition in this compound. Our investigations reveal glass-like arrest of kinetics at low temperature which plays a dominant role in the anomalous thermomagnetic irreversibility observed in this system. The interplay between kinetic arrest and supercooling is investigated by following novel paths in the H-T space. It is shown that coexisting FMM and AFI phases can be tuned in a number of ways at low temperature. These measurements also show that kinetic arrest temperature and supercooling temperature are anti-correlated i.e. regions which are arrested at low temperature have higher supercooling temperature and vice versa.
\end{abstract}
\pacs{75.47.Lx, 75.30.Kz}
\submitto{\JPCM}
\maketitle
 
\section*{Introduction}
The compound $Nd_{0.5}Sr_{0.5}MnO_{3}$ (NSMO) undergoes a paramagnetic insulator (PMI) to ferromagnetic metal (FMM) transition at ${T_C} \approx 255 K$ and FMM to antiferromagnetic insulator (AFI) transition at lower temperature ${T_N} \approx 150 K$ \cite{Kuwa,Toku}. The PMI to FMM transition is second order in nature, whereas the FMM to AFI transition is first order and shows a hysteretic temperature dependence of both resistivity and magnetization across the transition. Below $T_N$ this system showed large change in resistivity with the application of magnetic field which, was taken as the signature of the isothermal magnetic field induced transition from AFI to FMM state. The reverse FMM to AFI transition was observed with decreasing magnetic field and showed a hysteresis in the isothermal field dependence of resistivity (RH) \cite{Kuwa,Toku}. The hysteretic behaviour for field induced AFI to FMM transition in this compound and related systems \cite{Kuwa,Toku, Tok, Mori} is taken as a characteristic of a first order transition, where both the phases co-exist as (meta)stable phase between the upper critical field (field required for AFI to FMM transition) and lower critical field (field required for FMM to AFI transition). For a canonical first order transition, below  supercooling limit $T^*$, the initial state is recovered after going through a full cycle of the control variable (here magnetic field H). Such canonical behaviour is observed at 60 K for NSMO, see figure 2(a) of \cite{Kuwa}. However, the hysteretic region in RH expands with decreasing temperature and finally showed an open hysteresis loop at low temperatures as shown in figure 2(b)-(d) of \cite{Kuwa},  which shows that system dose not recover its initial state even when the magnetic field cycle is completed and magnetic field is reduced back to zero. Also, the HT phase diagram \cite{Kuwa,Toku} based on isothermal RH measurements showed that upper critical field increases monotonically with decreasing temperature whereas the lower critical field varies non-monotonically with temperature (see figure 3 and 4 of \cite{Kuwa} and figure 3 of \cite{Toku}). Taking lower critical field as following the limit of supercooling curve raises difficulties discussed below as non-monotonic variation is not expected for supercooling limit in a conventional first order transition. The anomalous open hysteresis loop at low temperature shall be addressed in this paper.

We assert that a non-monotonic behaviour of supercooling limit is anomalous and demonstrate the problem using a schematic HT phase diagram as shown in Figure 1. This phase diagram is similar to that shown in earlier work \cite{Kuwa,Toku} except that we have interchanged the horizontal and vertical axis, and highlights the non monotonic variation of supercooling spinodal ($H^*,T^*$) or lower critical field. To understand the problem we consider that the system is cooled in constant magnetic field along path AB. Point A lies above the superheating spinodal ($H^{**},T^{**}$) at which the AFI state is unstable and FMM state is stable; the Landau free energy has single minimum for the FMM state. Point P lies on the super-heating spinodal and a local minimum corresponding to the AFI state is in the process of developing; there is an inflection point but no local minimum \cite{Chai}. Point Q shows both states correspond to local minima in the free energy and since the minima shown are equal it corresponds to the phase transition temperature. As we move from Q to R the minimum corresponding to FMM becomes shallower turning into an inflection point at R which sits on the supercooling spinodal and infinitesimally small fluctuation at point R will transform FMM phase to AFI phase. At lower temperatures like that depicted by point S there is only one minimum corresponding to AFI state. If the supercooling spinodal is non-monotonic then at point B the local minimum corresponding to FMM state must reappear. This is not in accordance with conventional wisdom of how the free energy density evolves with temperature \cite{Chai}.
\begin{figure}[t]
	\begin{center}
	\includegraphics{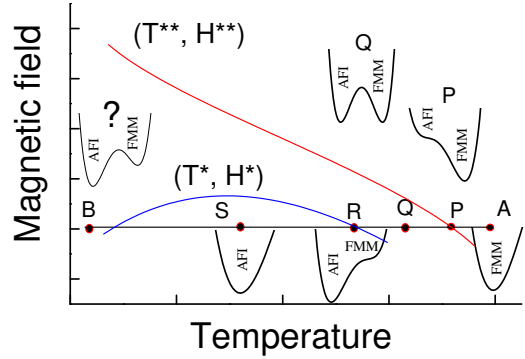}
	\end{center}
	\caption{Schematic HT diagram showing supercooling ($H^*, T^*$) and superheating ($H^{**}, T^{**}$) spinodal and Landau free energy curve for various points along path AB. The vanishing of energy barrier at point R and reappearance at point B is highlighted. See text for details.}
	\label{Figure 1}
\end{figure}

These arguments show that anomalous behavior in NSMO can not be explained by supercooling alone or on the basis of conventional first order phase transition. To propose an alternative to earlier experimental phase diagram \cite{Kuwa,Toku}, one needs to go beyond the first order phase transition and address the issues related to the anomalies observed at low temperature. In fact such anomalous thermomagnetic irreversibility has been observed in a wide variety of systems  and has been extensively investigated during last few years \cite{Bane, Ban, Kran, Ghiv, Shar, Wu, Mane,  Mage, Chat, Kaus}. There is growing evidence that such irreversibility is a result of kinetics of first order transition getting hindered. Manekar et al. \cite{Mane} in their extensive studies of Al doped $CeFe_2$ had observed various anomalous behaviour including open hysteresis loop similar to that observed in NSMO \cite{Kuwa,Toku}. They have proposed interplay between kinetic arrest and supercooling which was tested by their extensive data on Al doped $CeFe_2$. They have suggested that such an interplay would explain the limited data available in NSMO also.

Pursuing this, we invoke kinetic arrest in the compound NSMO to explain observed anomalous behaviour at low temperatures. In this paper, we present a  systematic investigation of first order transition in NSMO, using detailed resistivity and magnetization measurements. In the following, we will first highlight various anomalous features associated with the first order transition which can not be explained by supercooling alone. Then we will show that metastable FMM phase observed at low temperature is a result of kinetically slow dynamics of first order transition rather than a supercooled metastable state. Based on these data we present a phase diagram incorporating kinetic arrest and supercooling/ superheating spinodal. We go beyond earlier work \cite{Kuwa, Toku, Mane} mainly in two accounts; First we test the validity of our proposal for NSMO by traversing novel paths in HT plane which were not traversed in earlier work. It will be shown that by following different measurement protocols one can control or tune the ratio of coexisting FMM and AFI phases at low temperatures. This tunability of coexisting FMM and AFI phases provides a control over their functional properties. Second and more significantly we correlate the disorder broadening of kinetic arrest band (${H_K,T_K}$) and supercooling band (${H^*,T^*}$) as has been proposed recently \cite{Chad}. Our results on NSMO reinforce the inference on other materials \cite{Kran} that regions which have lower supercooling spinodal temperature have their kinetics arrested at higher temperature.

\section*{Experimental Details}
Polycrystalline NSMO is prepared by solid state reaction method and characterized by powder x-ray diffraction. Reitveld analysis of x-ray diffraction data and iodometric titration show that the compound is single phase and stoichiometric. Resistivity measurements are performed by standard four probe technique using a commercial cryostat (Oxford Instruments Inc., UK) with 8-Tesla magnetic field.  Magnetization measurements are carried out using a commercial 14 Tesla Vibrating Sample Magnetometer (Quantum Design, PPMS-VSM). 

\section*{Results and discussion}
Figure 2(a) and (b), show the temperature dependence of the resistivity in the absence of magnetic field and magnetization in the presence of low field (500 Oe), respectively, during cooling and heating for the compound NSMO. Both the measurements show two transitions. The high temperature transition from paramagnetic insulator (PMI) to ferromagnetic metallic (FMM) phase is indicated by onset of a sharp increase in magnetization around 230 K and changeover from insulator like to metal like  behavior in resistivity around 200 K. At low temperatures a sharp rise in resistivity around 120 K and corresponding decrease in magnetization with decreasing temperature, indicates another transition from FMM to antiferromagnetic insulating (AFI) phase. Transition from FMM to AFI phase is accompanied by a large thermal hysteresis which indicates first order nature of this transition. As can be seen from Figure 2, the transitions during both cooling as well as warming, are quite broad ($\Delta T \approx 50 K$). The broad transition can be associated with disorder which is natural to multicomponent systems. Such disorder gives rise to a landscape of free energies resulting in a spatial distribution of phase transition line ($H_C, T_C$) across the sample \cite{Imry}. Therefore for a macroscopic system large number of ($H_C, T_C$) lines would form a band. Similarly the spinodal lines corresponding to the limit of supercooling (T*) and superheating (T**) would also be broadened into a band \cite{Kran, Mane, Chad}. Each of these bands are made of quasi-continuum of lines in which each line represents a region of the sample.  Present observations on our polycrystalline NSMO are in qualitative agreement with studies on single crystalline NSMO by Kuwahara et al. \cite{Kuwa} and Tokura et al. \cite{Toku}. They \cite{Kuwa,Toku} also observed a second order PMI to FMM transition at higher temperatures followed by a broad first order FMM to AFI transition at low temperature. The quantitative difference like relatively smaller change in resistivity could be due to disorder inherent in polycrystalline samples. However, these differences between the two samples are unimportant for the present discussion and conclusions drawn in this manuscript will be applicable to both kinds of samples.

\begin{figure}[h]
	\begin{center}
	\includegraphics{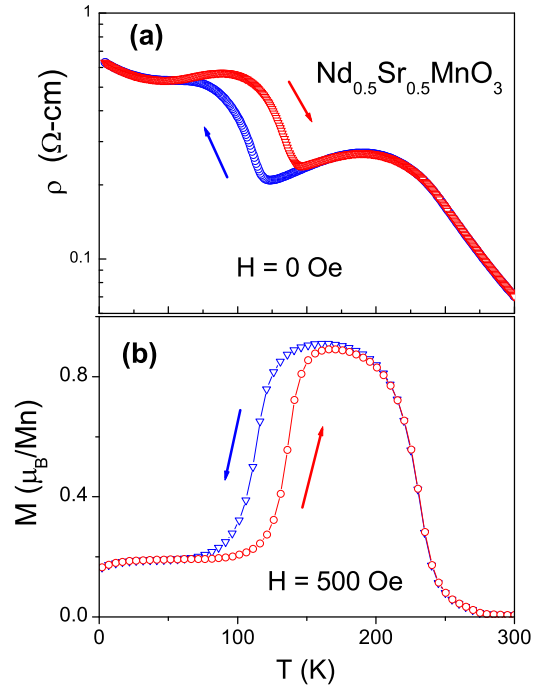}
	\end{center}
	\caption{Temperature dependence of (a) resistivity in the absence of magnetic field and (b) magnetization in the presence of 500 Oe magnetic field for the compound $Nd_{0.5}Sr_{0.5}MnO_{3}$ during cooling and warming run.}
	\label{Figure 2}
\end{figure}

The first order transition in this compound can be influenced dramatically with the application of magnetic field and shows strong history or path dependence. This can be seen from the temperature dependence of resistivity and magnetization in presence of various constant magnetic fields. Results for 1  Tesla and 2 Tesla magnetic field are shown in Figure 3. 
\begin{figure}[t]
	\begin{center}
	\includegraphics{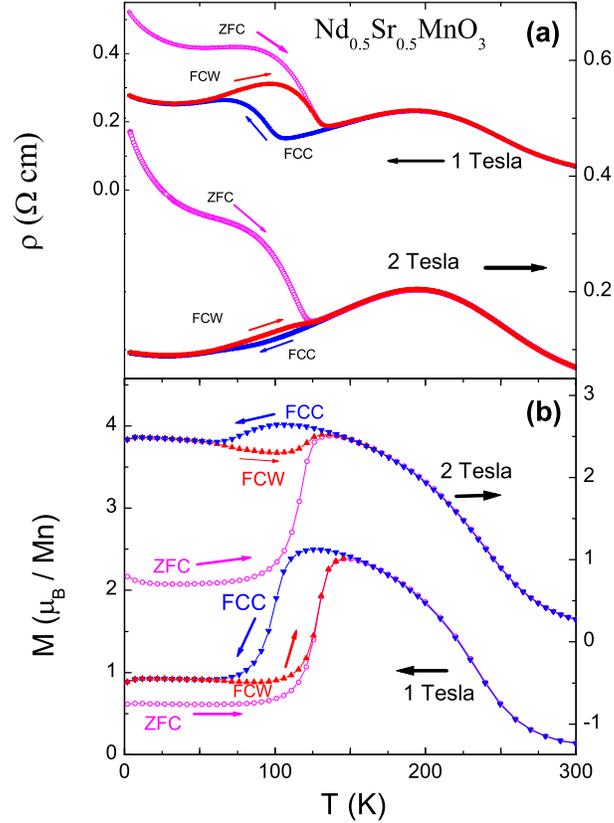}
	\end{center}
	\caption{Temperature dependence of (a) resistivity and (b) magnetization at 1 and 2 Tesla measured under different protocols. ZFC indicates measurement taken during warming for the labeled field applied at the lowest temperature after cooling the sample in zero field. FCC/FCW indicates measurement taken subsequently during cooling/warming with the field kept on. Large difference between ZFC (open symbol) and FCC/FCW (solid symbol) is clearly visible in all the measurement and ZFC is laying outside FCW and FCC curves.}
	\label{Figure 3}
\end{figure}
For these measurements samples were cooled under zero field condition to the lowest temperature of measurement and then above mentioned field applied isothermally at this temperature. The resistivity and magnetization is measured under above mentioned constant magnetic field in following sequence: (a) warming to highest temperature of measurement (ZFC), (b) cooling under same field to lowest temperature (FCC) and again (c) warming under same field (FCW). As can be seen from Figure 3, the transition from FMM to AFI state shifts to lower temperature at higher magnetic field. These observations are consistent with the resistivity measurement on single crystalline NSMO \cite{Toku}. However, ZFC curves, which were not shown earlier for NSMO, shows anomalous behavior. At low temperatures  there is a large difference between ZFC and corresponding FC (field cooled) curves. Similar behavior is observed in magnetization measurements (Fig. 3(b)), which shows distinctly higher magnetization for FCW/FCC curve compared to ZFC below transition temperature. The difference between ZFC and FC curve persist up to or above 100 K. Another interesting feature is that difference between the ZFC and FC curves increases with increasing magnetic field. Though such bifurcation in the history dependent magnetization is expected to arise for metastable magnetic systems or domain wall dynamics in ferromagnets, contrary to present observation (Figure 3), it is generally not coupled with similar bifurcation in resistivity. Moreover, the increase in the difference between the ZFC and FC curves with increasing magnetic field dose not take place for any of the above mentioned systems contrary to the observation in Figure 3(b). For all such conventional systems the bifurcation between ZFC and FC magnetization actually decreases with the increase in field. Therefore different behavior for FC and ZFC curve indicates different magnetic states across these paths. Since FC curve have higher magnetization and lower resistivity we can infer that FC curve corresponds to larger fraction of FMM phase compared to ZFC curve.

Another anomalous feature associated with this transition can be seen in isothermal magnetization measurements. Figure 4 (a) and (b) show MH curves recorded at various constant temperatures. For these measurements sample was cooled under zero field condition from room temperature to measurement temperature before recording the MH curve. 
\begin{figure}[b]
	\begin{center}
	\includegraphics{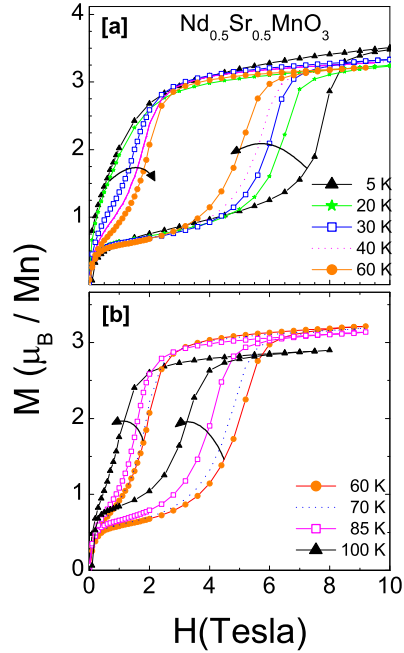}
	\end{center}
	\caption{The magnetic field dependence of magnetization at various temperatures showing return MH move (a) to higher field values below 60K and (b) to lower field values above 60K.  Whereas the forward MH curve moves to lower field values monotonically with increasing temperature}
	\label{Figure 4}
\end{figure}
The forward MH curve ($0 \to H_{max}$) shows a sharp step in magnetization with increasing magnetic field, indicating a field induced AFI to FMM transition. As can be observed from Figure 4(a) and (b) the forward MH curve ($0 \to H_{max}$) moves monotonically to lower field values with increasing temperature from 5 to 100 K. It indicates that the magnetic field required for the AFI to FMM transition becomes smaller monotonically with increasing temperature. Similarly the return MH curve ($H_{max} \to 0$), which reflects the transition from FMM to AFI state, moves to lower field values with increasing temperature above 60 K (Figure 4(b)). This is expected as the FMM phase is stable to lower temperature in higher magnetic field. In contrast to this,  below 60 K, return MH curves (Figure 4(a)) show opposite trend i.e. transition from FMM to AFI shifts to lower field values with decreasing temperature. This indicates that below 60 K FMM phase exists down to lower field values at lower temperature. For a conventional first order transition such non monotonic behavior is not expected. This anomaly is demonstrated more explicitly in the HT phase diagram.

Figure 5 shows this phase diagram for first order transformation in NSMO, which is obtained from isothermal MH (as shown above) and RH curves. This phase diagram shows the temperature dependence of $H_{up}$ (critical field required for AFM to FMM transition) and $H_{dn}$ (critical field required for FMM to AFM transition). These values of critical magnetic field are taken as the magnetic field at which first derivative of magnetization in MH and second derivative of resistivity in RH shows an extremum. Due to absence of sharp features in RH curves across the transition, calculated critical field can have larger error compared to that obtained from MH curves. This phase diagram shown in Figure 5 is qualitatively similar to phase diagram of single crystalline NSMO \cite{Kuwa,Toku}. As expected, $H_{up}$ increases with decreasing temperature as AFI phase is stable at low temperature. Whereas $H_{dn}$ initially increases with decreasing temperature, shows a shallow maximum around 60 K ($T_a$) and then starts decreasing with further lowering of temperature. The consequences of such non-monotonic variation of $H_{dn}$ are discussed in introduction section where it is shown that supercooling alone is not sufficient to explain such behaviour. 

\begin{figure}[ht]
	\begin{center}
	\includegraphics{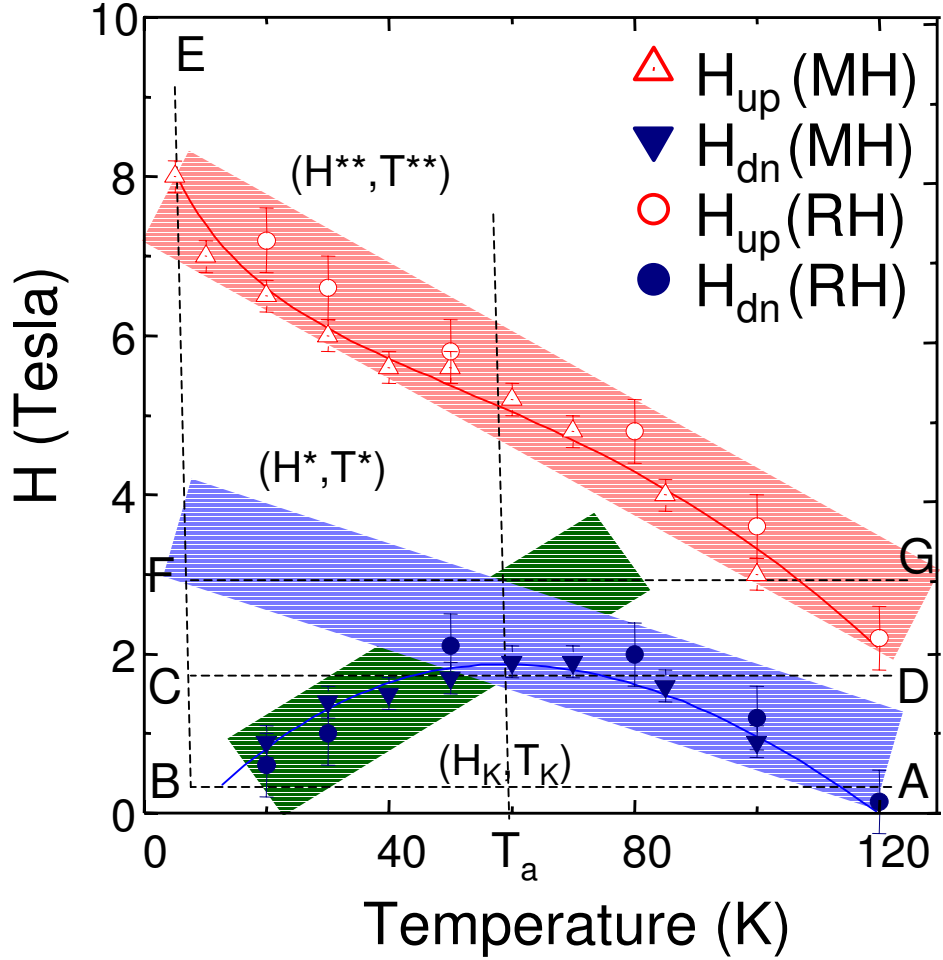}
	\end{center}
	\caption{(a) HT diagram for the compound $Nd_{0.5}Sr_{0.5}MnO_{3}$ (NSMO) showing critical field required for AFI to FMM transitions (Hup) and FMM to AFI transition (Hdn). Triangles and circles respectively represent the critical field obtained from M vs. H curve and R vs. H curves.  Shaded regions shows the kinetic arrest $(H_K,T_K)$, supercooling ($H^*, T^*$) and superheating ($H^{**}, T^{**}$) spinodals for NSMO.}
	\label{Figure 5}
\end{figure}

Some of these anomalous behaviour observed for NSMO, had also been observed in other system e.g. $Ce(Fe_{0.96}Al_{0.04})_2$ \cite{Mane,Sing}, Pr doped $La_{5/8}Ca_{3/8}MnO_{3}$ \cite{Ghiv, Shar}, $Pr_{0.5}Ca_{0.5}Mn_{0.975}Al_{0.025}O_{3}$ \cite{Ban}, $Pr_{0.5}Sr_{0.5}MnO_{3}$ \cite{Bane} etc. In these studies such behaviour is attributed to arrested metastable high temperature phase coexisting with stable phase at low temperatures. In case of $Ce(Fe_{0.96}Al_{0.04})_2$ \cite{Mane,Sing} it has been shown that due to critically slow dynamics of phase transition, the transformation from high temperature to low temperature phase is hindered on measurement time scale, resulting in the  glass like arrest of a long range ordered state at low temperatures \cite{Chatt}. It is worth to note here that such kinetically arrested metastable (glass) states are different from the metastable supercooled state \cite{Kran,Chad}. It has been already highlighted in Kranti Kumar et al \cite{Kran}, that supercooled state (but not the metastable kinetically arrested state) will undergo metastable to stable transformation on lowering temperature. In other words relaxation time decreases with decrease in temperature for supercooled state whereas the relaxation time increases with decrease in temperature for glassy state. 

Therefore the metastable FMM state at low temperature in NSMO can be treated as kinetically arrested state.  For this we consider ($H_K,T_K$) band below which transformation from high temperature phase (FMM) to low temperature phase (AFI) is hindered on measurement time scale. In other words, if a FMM phase crosses the kinetic arrest band ($H_K,T_K$) before crossing the corresponding supercooling spinodal then it will remain in FMM phase even after crossing the supercooling spinodal ($H^*,T^*$). These bands i.e. kinetic arrest band ($H_K,T_K$), supercooling band ($H^{*},T^{*}$) and superheating band ($H^{**},T^{**}$) are shown in Figure 5 as hatched area. As argued previously \cite{Imry}, these bands are constituted of the representative ($H_K,T_K$), ($H^*,T^*$) and ($H^{**},T^{**}$) lines related to different regions of the samples. The anomalous features observed in Figure 3 and 4, now can be explained using this new HT diagram. For example in Figure 3, measurements were performed along path CD (Figure 5) after cooling the sample following different paths. For ZFC curve, we reached the lowest temperature i.e. point C, along path AB (under zero field) and then applied magnetic field isothermally to reach point C. Since along path AB supercooling band ($H^*,T^*$) is traversed before crossing the  arrest band ($H_K,T_K$), sample will be completely in AFI state at point B. Sample will remain in AFI state at point C as it is below ($H^{**},T^{**}$) band. So ($H_K,T_K$) band has no role to play. Therefore ZFC curve along path CD will start with completely AFI state.  The transition to FMM phase will occur on crossing the ($H^{**},T^{**}$) band. FCC curve will starts from point D with fully FMM phase and transformation to AFI phase will complete on crossing the ($H^*,T^*$) band. However, for some regions of the sample the ($H_K,T_K$) lines will be crossed before the corresponding ($H^*,T^*$) lines \cite{Kran}. Therefore, there will be only a partial transformation from FMM to AFI phase and some regions will get arrested in FMM state. Since FMM phase is arrested on cooling and cannot transform to the equilibrium AFI phase on measurement time scale, this coexistence will persist in field cooling down to the lowest temperature. This is in contrast to ZFC curve, where at point C entire sample is in AFI state therefore having much higher resistivity or lower magnetization compared to FCC. And now, FCW curve will start with partial frozen FMM phase. On warming, the dearrest of FMM phase will start on crossing the ($H_K,T_K$) band in opposite sense. This is evident from the decrease in M or increase in R of the FCW path after it bifurcate from FCC curve at low temperature (figure 3). Subsequent warming will lead to the transformation of AFI phase to FMM phase on crossing the ($H^{**},T^{**}$) band. If we increase the magnetic field to say 2 Tesla, then arrest band will appear at higher temperature resulting in more arrested FMM phase on field cooling. Beyond certain field values entire sample will cross ($H_K,T_K$) band before crossing the ($H^*,T^*$) band during field cooling (e.g. path GH) resulting in a completely arrested FMM phase at low temperature.

In above experiments the arrested FMM phase is obtained by varying temperature in presence of constant magnetic field i.e. when ($H_K,T_K$) band is crossed before ($H^*,T^*$) band during field cooling. Sample cooled in zero field to the left of ($H_K,T_K$) band is in completely AFI phase (Figure 5). This stable AFI phase can be converted to kinetically arrested FMM phase by varying magnetic field at low temperature. When we apply a magnetic field large than $H^{**}$ on a zero field cooled sample, say along path BE, it is transformed completely from AFI state to FMM state. Now reducing magnetic field to zero will not result in AFI state until and unless sample crosses the ($H_K,T_K$) band in opposite sense.  Therefore we can have different magnetic states at H=0 before and after the application of magnetic field. Before the application of magnetic field we had stable AFI phase whereas after the application of magnetic field we get kinetically arrested FMM phase along with AFI phase. Further field cycling will transform the remaining AFI phase to and from FMM phase whereas kinetically arrested FMM phase will remain frozen. Some of such isothermal MH and RH loop measurements are shown in Figure 6 for NSMO. For these isothermal measurements sample is cooled under zero field to measurement temperature and then magnetic field is applied to measure the field dependence of resistivity and magnetization. With increasing magnetic field (path i: $0 \to H_{max}$) both resistivity and magnetization  shows magnetic field induced transition from AFI to FMM phase. With decreasing field (path ii: $H_{max} \to 0$) the reverse transition from FMM to AFI as seen in Figure 6, starts at lower field compared to $H_{up}$ and finally at low temperatures FMM to AFI transformation does not complete even down to zero magnetic field which is clearly visible in RH data shown in Figure 6(a) and (b). These curves shows two distinct anomalies: (i) There is a large difference between zero field resistivity before and after the application of magnetic field at low temperature. This observation is similar to that observed by Kuwahara et al. and Tokura et al. in their single crystalline NSMO \cite{Kuwa, Toku, Tok}. (ii) Secondly, the virgin curve (path i: $0 \to H_{max}$) is lying outside the envelope curve (path iii to v: $0 \to -H_{max} \to +H_{max}$) for both MH and RH measurements. This anomaly does not show up unless the path (iii to v) is also traversed. Both the anomalies becomes pronounced at low temperatures. These behaviours arise due to arrested FMM phase obtained after the application of magnetic field. Since only a fraction of sample crosses ($H_K,T_K$) band at the end of path (ii) along EB, we get the coexisting stable AFI phase along with kinetically arrested metastable FMM phase at zero field. Therefore zero field resistivity is smaller after the application of magnetic field. With increasing temperature more and more fraction crosses ($H_K,T_K$) band with decreasing magnetic field which results in an increased AFI fraction at the end of path (ii). 
\begin{figure}[t]
	\begin{center}
	\includegraphics{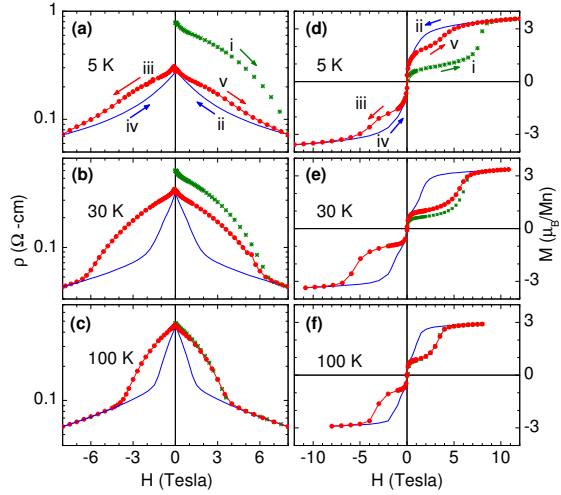}
	\end{center}
	\caption{Magnetic field dependence of resistivity and magnetization at various constant temperatures. A large difference between zero field resistivity before and after the application of magnetic field and virgin curve (open symbols) lying outside the envelop curve (solid symbol) at low temperatures. See text for details.}
	\label{Figure 6}
\end{figure} 
Therefore the difference between zero field resistivity before and after the application of magnetic field decreases at higher temperature. This explains our data in figure 6 and also the measurements of Kuwahara et al. \cite{Kuwa}.

The envelope curve at low temperature starts with kinetically arrested metastable FMM phase and stable AFI phase and reflects the transformation of AFI phase along with field dependence of FMM phase. Therefore virgin curve will lie outside the envelope curve as the arrested FMM phase has higher magnetization (lower resistivity). The virgin and envelope curves merge at high field where the sample is in homogeneous FMM state. The difference between envelope and virgin curve will decrease with increasing temperature due to decrease in arrested FMM fraction at the end of path (ii). For temperatures $T \geq T_a$, ($H^*,T^*$) band will determine the FMM to AFI transition during isothermal measurements, as with decreasing magnetic field de-arrest will take place before hitting the corresponding supercooling line for a particular region. 
 
These observations show that the transformation from FMM to AFI phase is kinetically arrested below ($H_K,T_K$). In other words if there is any FMM phase to the left of ($H_K,T_K$) band then it will not transform back to AFI phase until and unless it crosses the ($H_K,T_K$) band in opposite sense. Therefore we expect arrested FMM phase to convert to AFI phase with reducing field after crossing the ($H_K,T_K$) band. These tendencies were probed by field annealing measurement. These measurements not only test the validity of our phase diagram but also brings in to light new observation. For these measurements a constant magnetic field $H_{an}$ (annealing field) is applied well above ($H_C,T_C$)  and then sample is cooled to selected temperatures to record isothermal RH or MH curve with reducing field from $H_{an}$ to zero. Results of such measurement are shown in figure 7, which highlights following features; 

\begin{figure}[t]
	\begin{center}
	\includegraphics[width=7cm]{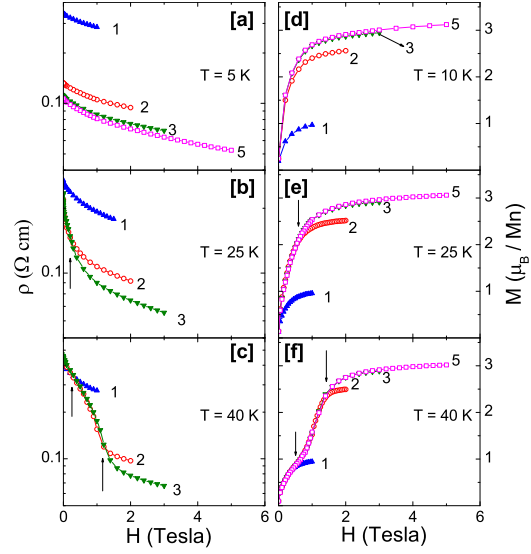}
	\end{center}
	\caption{((a)-(c) RH and (d)-(e) MH at various constant temperatures for NSMO cooled under different annealing field $H_{an}$ from $T>T_N$. Vertical arrows indicate dearrest field. These measurements across novel path test the correlation between kinetic arrest and supercooling band. See text for details.}
	\label{Figure 7}
\end{figure} 

(i) Both resistivity and magnetization depends on annealing field $H_{an}$. 

(ii) For $H_{an}\leq 3$ Tesla, both RH and MH curves are almost distinct. This feature is more prominent at low temperature where a large difference is observed even when magnetic field is reduced to zero, e.g. see figure 7(a). 

(iii) The RH and MH curve merge together with decreasing magnetic field at 25 K and 40 K. The curves for higher $H_{an}$ merge in the lower temperature measurements (contrast figure 7(b)and 7(e) with 7(c) and (f)).

(iv) The magnetic field at which these curves merge together increases with increasing temperature. The merger points are shown by vertical arrow in figure 7(b),(c),(e) and (f). The significance of these features are discussed below.

At 5 K, as shown in Figure 7(a), resistivity values are dependent on annealing fields $H_{an}$ in which sample is cooled to 5 K. Smaller resistivity values with increasing $H_{an}$ indicate larger kinetically arrested FMM phase as with increasing $H_{an}$ more and more regions cross corresponding kinetic arrest line before crossing the supercooling line during cooling. For magnetic field 5 and 3 Tesla, resistivity curves are almost identical, which indicates for $H_{an}\geq 3$ Tesla entire sample crosses the ($H_K,T_K$) band before crossing the ($H^*,T^*$) band. This is consistent with magnetization at 10K (Figure 7(d)) where for fields higher than 3 Tesla magnetization has negligible dependence on $H_{an}$ and it is around $3\mu_B$/Mn. From a comparison with saturation magnetization at 10 K we infer almost fully ferromagnetic phase  for  annealing field larger than 3 Tesla. For annealing field 2 and 1 Tesla magnetization is equivalent to $\approx 80\%$ and $30\%$ of FMM phase, respectively. 

The observed change in resistivity with reducing magnetic field at 5 K is much smaller compared to resistivity difference produced by different annealing field. RH curves remains distinct with reducing magnetic field down to zero. It results in a large difference in resistivity values even at H=0 for different annealing field. This shows that with reducing magnetic field at 5 K only a small fraction of sample crosses the ($H_K,T_K$) band as magnetic field is reduced to zero resulting in large difference between zero field resistivity values. Therefore most of the arrested FMM phase remains frozen down to zero magnetic field (This feature is not so obvious in MH measurement, see Figure 7(d), as magnetization goes to zero for FMM phase also at zero field). Therefore field annealing measurements indicate that one can tune the FMM fraction at low temperature by varying annealing field.

At 25 K and 40 K also, resistivity and magnetization both are dependent on $H_{an}$ as shown in Figure 7(b),(d), (e) and (f). This is expected if these temperature and field values are to the left of ($H_K,T_K$) band in figure 5. However, in contrast to 5 K, at 25 K both RH and MH curves for $H_{an}$ = 2 and 3 Tesla merges around $\approx 0.5$ Tesla indicated by vertical arrow in the figure 7(b) and (e). Whereas 1 Tesla curve remains distinct down to 0 Tesla. It suggests that regions which are arrested for field $H_{an} \geq 2$ Tesla show de-arrest at 25 K with reducing magnetic field. At 40K, as shown in figure 7(c) and (f),  all the curves ($H_{an}$ = 1, 2 and 3 Tesla) merge together with decreasing magnetic field which indicates  that regions which are arrested for field values $H_{an} \geq 1$ Tesla are de-arrested with decreasing magnetic field. Also the curve for $H_{an}$ = 2 and 3 Tesla merge together before merging with 1 Tesla curve with decreasing magnetic field. Secondly the merger point for  $H_{an}$ = 2 and 3 Tesla curve is shifted to higher field values at 40 K as compared to 25 K. These results show that region arrested in higher annealing field lie to the high temperature end of the ($H^*,T^*$) band, but de-arrest at higher H on isothermal reduction of field indicates such regions lie to the low temperature end of the ($H_K,T_K$) band. In fact the behavior of MH curves shown in Figure 7(f) is similar to schematic diagram shown in figure 2 (b) of Chaddah et al. \cite{Chad}, where correlation between kinetic arrest band and supercooling spinodal has been addressed. It was shown that such kind of MH behaviour is expected for anticorrelated kinetic arrest band and supercooling spinodal. Similar MH behaviour is also observed for the compound $Ce(Fe_{0.98}Os_{0.02})_2$ \cite{Kran}, which also shows a first order FM to AFM transition with lowering temperature and kinetically arrested ferromagnetic phase. From these studies we infer that ($H_K,T_K$) band and ($H^*,T^*$) band are anticorrelated in NSMO i.e. regions which have higher ($H_K,T_K$) have lower ($H^*,T^*$). It is interesting to note here that so far anticorrelation appears to be universal feature.  Disorder seems to hinder the kinetics of first order transition. Therefore it results in a higher kinetic arrest temperature for a more disordered region whereas the transition temperature decreases with increasing disorder. Such studies can provide more insight to  glassy transitions due to easy tunability of magnetic field.

In this paper, we have not reported measurements of slow relaxation in magnetization and in resistivity. The data shown corresponding to the kinetic arrest band in figure 5 is obtained when rapid de-arrest of the FMM phase takes place. The unstable but kinetically arrested FMM phase transforms rapidly to the stable AFI state either on raising temperature or on lowering H, akin to the de-vitrification of glass. The high-field end of this band lies above the supercooling band, and that can only be mapped by measurements of slow relaxation where the relaxation time will diverge as one lowers temperature and approaches T$_{K}$ \cite{Chatt}. Such studies will be undertaken in future work.

\section*{Conclusions}
Anomalous nature of first order AFI to FMM transition in NSMO has been elucidated by transport and magnetic measurements. It has been shown that one has to invoke glass like arrest of kinetics to explain observed thermo-magnetic irreversibility in this system. This also explains that the FMM to AFI transition (de-arrest) with magnetic field  is broadened as temperature is lowered while the AFI to FMM transition becomes sharper (see Figure 2 of \cite{Kuwa}). A H-T phase diagram consisting of kinetic arrest band and supercooling and superheating spinodal has been proposed for this system. It has been shown that below ~60 K, kinetic arrest dominates the FMM to AFI transition. This results in coexistence of AFI and kinetically arrested FMM phase at low temperature, even though the equilibrium phase is AFI. The ratio of these coexisting phases can be controlled by traversing novel paths in HT space. The competing tendencies of kinetic arrest and supercooling is studied by field annealing measurements. It has been shown that for $H_{an}\geq$ 3 tesla entire sample can be arrested in to FMM state at low temperature. These measurements also indicate that regions arrested at higher temperature ($H_K,T_K$) have lower supercooling temperature ($H^*,T^*$) and vice versa. This anticorrelation between kinetic arrest and supercooling seem to be a universal feature. Since this model depends on few parameters and describes the state of a system for any path in HT space, similar studies will be useful for several other such systems where blocked metastable states have been observed.    
 
\ack{Acknowledgments}
We would like to thank S.B. Roy (RRCAT, Indore) for fruitful discussions. DST, Government of India is acknowledged for funding the VSM. KM acknowledges CSIR, India.

\end{document}